\documentclass[copyright]{eptcs}
\providecommand{\event}{DCFS 2010} % Name of the event you are submitting to
\usepackage{breakurl}             % Not needed if you use pdflatex only.

\usepackage{graphicx}
\usepackage{epstopdf}
\title{L-systems in Geometric Modeling}
\author{Przemyslaw Prusinkiewicz \qquad\qquad Mitra Shirmohammadi \qquad\qquad 
Faramarz Samavati
\institute{Department of Computer Science, University of Calgary,\\
2500 University Dr. N.W., Calgary, AB T2N 1N4, 
Canada}
\email{\quad pwp@cpsc.ucalgary.ca \quad\qquad mitrashm@yahoo.com \quad\qquad samavati@cpsc.ucalgary.ca}
}
\def\titlerunning{L-systems in Geometric Modeling}
\def\authorrunning{P.~Prusinkiewicz, M.~Shirmohammadi \& F.~Samavati}

\begin{document}

\maketitle

\begin{abstract}
We show that parametric context-sensitive L-systems with affine geometry interpretation provide a succinct description of some of the most fundamental algorithms of geometric modeling of curves.  Examples include the Lane-Riesenfeld algorithm for generating B-splines, the de Casteljau algorithm for generating B\'ezier curves, and their extensions to rational curves. Our results generalize the previously reported geometric-modeling applications of L-systems, which were limited to subdivision curves.
\end{abstract} 

\section{Introduction}
\label{sec:Introduction}

L-systems were conceived by Aristid Lindenmayer as a mathematical formalism for reasoning about growing multicellular organisms~\cite{Lindenmayer1968}.  They were originally introduced as an extension of cellular automata, allowing for the addition and removal of cells during an automaton's operation, but were soon rephrased in terms of rewriting systems~\cite{Lindenmayer1971}.  This rephrasing led to an elegant definition of L-systems using notions and notation borrowed from formal language theory.  The relation to formal languages colored early studies of L-systems, which were often focused on their ability to generate different classes of languages~\cite{Herman1975,Rozenberg1980}.  

Concurrent with these developments, although initially on a smaller scale, L-systems began to be used as a mathematical foundation for the computational modeling of plants.  The fundamental observation was that a growing filament, i.e., a linear arrangement of cells, can be conveniently viewed as a word over an alphabet of cell states.  This abstraction was generalized to branching structures and components larger than individual cells, leading to astonishingly succinct descriptions of growing plants~\cite{Prusinkiewicz1990a}.  

In retrospect, the success of L-systems in plant modeling can be attributed to the following factors~\cite{Prusinkiewicz2009dc}:
\begin{enumerate}
\item  
L-systems specify development in terms of temporally and spatially local declarative rewriting rules --- context-free and context-sensitive productions --- which express developmental processes in a concise and intuitive manner. 
\item
L-systems describe growing structures in terms of topological (neighborhood) relations between structure components, which are automatically maintained when components are added to or removed from a structure.  Consequently,  the context for production application is always available.
\item 
L-systems refer to model components by their type, state and context, rather than a unique name or position in the structure.  This simplifies the specification of developmental algorithms.  In particular, indices are no longer needed.
\end{enumerate}  

\noindent
These features are not only important to the modeling of biological development, but also make L-systems well suited to the description and implementation of some non-biological algorithms~\cite{Prusinkiewicz1992b}.  Typically, they are characterized by the repetitive application of relatively simple rules to discrete structures with a changing number of components.  Early studies of such algorithms led to the concise L-system description of linear fractals, including classic space-filling curves~\cite{Prusinkiewicz1986,Prusinkiewicz1991ssf,Siromoney1983,Szilard1979}.  The geometric interpretations of L-systems used in these studies were based on turtle geometry~\cite{Abelson1982} or chain coding~\cite{Freeman1961}. 

More recently, L-systems with an interpretation based on affine geometry~\cite{deRose1992tdc,DeRose1989acf,Goldman2002ota,Goldman2003pa} have been demonstrated to provide a concise description of the subdivision curves used in geometric modeling~\cite{Prusinkiewicz2002gmw,Prusinkiewicz2003lsd}.  Similar to the algorithms for generating fractal curves, subdivision algorithms operate on discrete, polygonal lines with a changing number and configuration of components. The rules of change can be conveniently described in local terms and formalized as L-systems productions. A comparison between L-systems and the traditional specification of subdivision algorithms using indexed points and/or matrices reveals the advantages of L-systems~\cite{Prusinkiewicz2003lsd}. 

Here we expand the range of geometric modeling applications of L-systems and show that L-systems allow for the succinct formulation of some of the most fundamental algorithms for geometric modeling of curves.  These include the Lane-Riesenfeld algorithm for generating B-splines, the de Casteljau algorithm for generating B\'ezier curves, and their extensions to rational curves.  Further examples are discussed in detail by Shirmohammadi~\cite{Shirmohammadi2008gmw}, whose work was a stepping stone for the present paper.  Compared to~\cite{Shirmohammadi2008gmw}, we use a topologically more accurate representation of polygons, rooted in the notion of a cell complex~\cite{Palmer1993cmo}, with both vertices and edges explicitly represented. This results in a more intuitive specifications of the algorithms, reflecting their inherent symmetries.   

\section{Preliminaries}

The L-systems employed here are context-sensitive parametric L-systems~\cite{Lindenmayer1974,Prusinkiewicz1990a}, extended in two directions:
\begin{itemize}
\item
Parameters are not limited to numbers, but may also be compound data structures, for example representing points or vectors in two or three dimensions~\cite{Karwowski2003dai,Prusinkiewicz2003lsd};
\item
Productions may be grouped into subsets (tables~\cite{Rozenberg1973tol}), with a control mechanism deciding which subset is applicable in each derivation step~\cite{Dassow1986ocl,Prusinkiewicz2007tlc,Yokomori1986gcs}. 
\end{itemize} 

The key concepts can be summarized as follows.  Parametric L-systems operate on parametric words, or strings of modules.  Each module is a letter from the L-system alphabet $V$, which may be associated with one or more optional parameters. Beginning with an explicitly defined axiom (initial string), an L-system generates a developmental sequence of words using a finite set of productions of the form:
\begin{equation}
label : \; \; left \; context < strict \; predecessor > right \; context : condition \rightarrow successor .
\end{equation}

\noindent
In each derivation step, productions are applied in parallel to all modules in the predecessor string.  If several productions apply, the first production in the list is chosen.  If no production applies, a module is rewritten into itself. For example, the L-system with axiom $\omega$ and production $p_1$ to $p_3$, given below
\begin{equation}
\begin{array}{ll}
\omega : & \lefteqn{A(1.5) B(2.0, 3.0) A(4.5) C(1)} \\
p_1 : & A(x)  : x \leq 2 \rightarrow A(2x+1) \\
p_2 : & A(x)  : x > 2 \rightarrow B(2x+1) \\
p_3 :& A(w) < B(x,y) > A(z) \rightarrow A(w+x) A(y+z)
\end{array}
\end{equation}
generates the word $A(4) A(3.5) A(7.5) B(10) C(1)$ in the first derivation step.  In addition to proper L-systems, in which the strict predecessor is always a single module, we consider pseudo-L-systems~\cite{Prusinkiewicz1986}, in which the strict predecessor may be a nonempty parametric word over the alphabet $V$.  We also consider L-systems operating on circular words, in which the first and last module are considered neighbors.

\begin{figure}
\begin{center}
\includegraphics[width=0.6\linewidth]{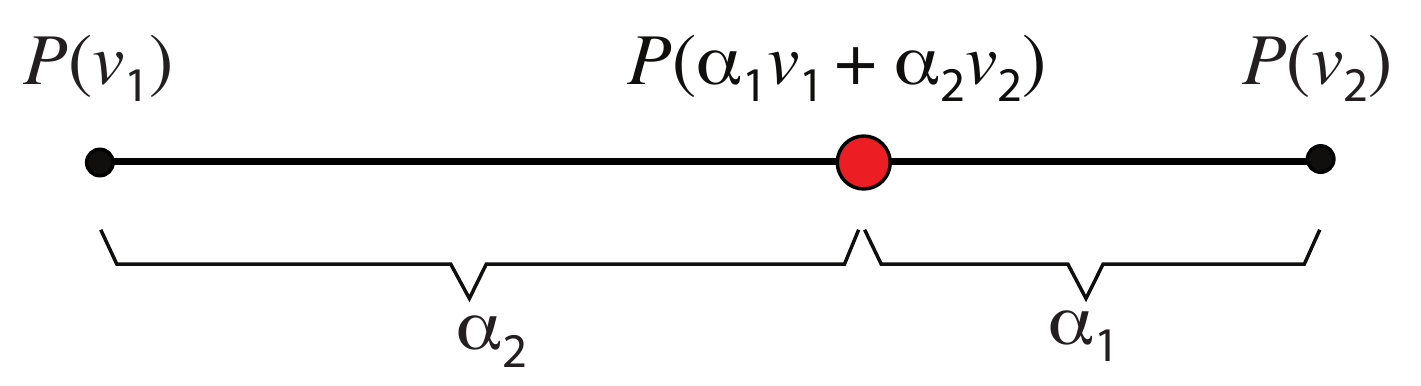}
\end{center}
\caption{Affine combination of points $P(v_1)$ and $P(v_2)$.}
\label{fig:affine-combination}
\end{figure}

Parameters may be $d$-tuples of numbers representing positions $v$ in a $d$-dimensional space~\cite{Prusinkiewicz2003lsd}.  An affine combination $v$ of $m$ positions
$v_1, v_2, \ldots, v_m$ is given by the expression
\begin{equation}
\label{eq:affine-comb}
v = \alpha_1 v_1 + \alpha_2 v_2 + \cdots + \alpha_m v_m ,
\end{equation}
where the scalar coefficients $\alpha_i$ add up to 1:
\begin{equation}
\alpha_1 + \alpha_2 + \cdots + \alpha_m = 1 .
\end{equation}
The meaning of the affine combination~(\ref{eq:affine-comb}) is derived
from its transformation to the form
\begin{equation}
\label{eq:affine-vector}
v = v_1 + \alpha_2 (v_2 - v_1) + \cdots + \alpha_n (v_m - v_1) ,
\end{equation}
which is a well-defined expression of vector algebra.  Specifically,
for two positions $v_1$ and $v_2$ we obtain:
\begin{equation}
v = \alpha_1 v_1 + \alpha_2 v_2 = v_1 + \alpha_2 (v_2 - v_1) =
v_2 + \alpha_1 (v_1 - v_2) ,
\end{equation}
which means that point $P$ with position $v$, noted $P(v)$, divides line $\overline{P(v_1) P(v_2)}$ in proportion $\alpha_2 : \alpha_1 $ (Figure~\ref{fig:affine-combination}).

\section{B-splines}

\begin{figure}
\hspace{1.6cm} a \hspace{3.8cm} b \hspace{3.9cm} c \hspace{3.5cm} d
\begin{center}
\vspace{-3mm}
\includegraphics[width=1\linewidth]{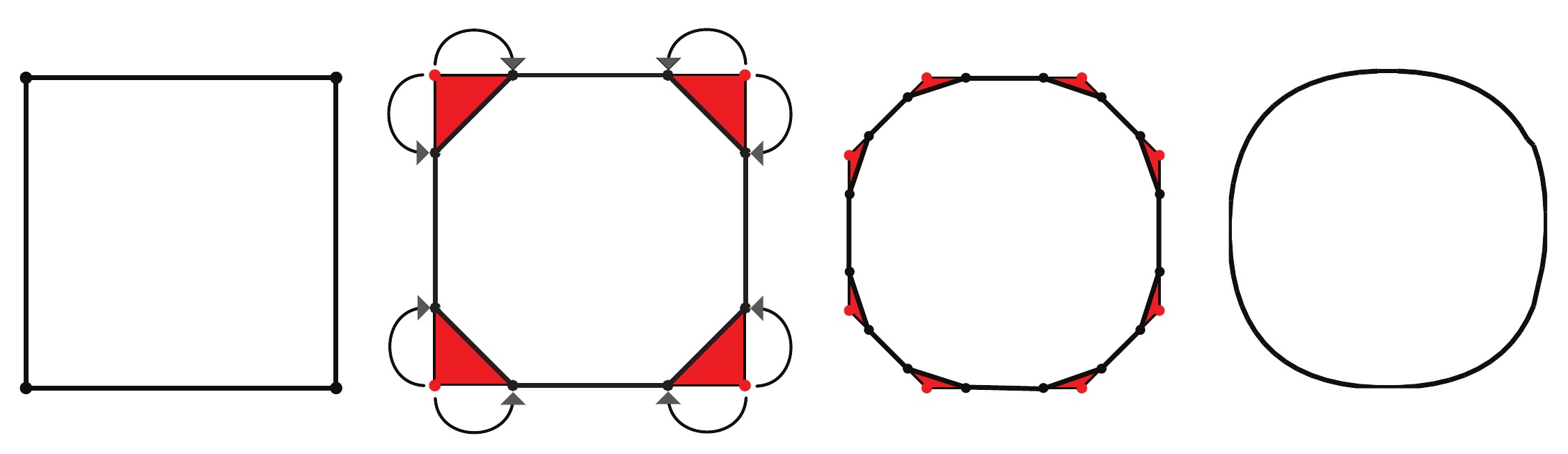}
\end{center}
\caption{Illustration of Chaikin's algorithm (adapted from~\protect\cite{Prusinkiewicz2003lsd}).  Beginning with a user-specified control polygon (a), the algorithm iteratively removes the polygon corners (red triangles in panels b and c).  Each old vertex is thus replaced by a pair of new vertices, moved $\frac{1}{4}$ of the distance toward the neighboring old vertices (arrows in panel b).  An approximately smooth curve obtained after 4 derivation steps is shown in panel d.}
\label{fig:Chaikin}
\end{figure}

A simple example of the application of L-systems to geometric modeling is an L-system specification of the Chaikin algorithm~\cite{Chaikin1974aaf}.  Given a (closed) control polygon $P(v_0)P(v_1) \ldots P(v_m)$, this algorithm produces a smooth (at the limit) curve, whose shape can be thought of as the result of iteratively cutting the corners of the control polygon and its descendants (Figure~\ref{fig:Chaikin}). This process can be succinctly specified by a context-sensitive L-system operating on circular words:
\begin{equation}
\label{eq:Chaikin}
\begin{array}{llcl}
\omega : & \lefteqn{P(v_1) P(v_2) \ldots  P(v_m)} \\
p: & P(v_l) < P(v) > P(v_r)  \rightarrow  
	P(\frac{1}{4} v_l + \frac{3}{4}v) P(\frac{3}{4} v + \frac{1}{4}v_r)
\end{array}
\end{equation}
It is often convenient to explicitly represent not only the vertices, but also the edges of the polygons on which the Chaikin algorithm operates.  Such a representation is provided by the following modification of L-system~\ref{eq:Chaikin}:
\begin{equation}
\label{eq:Chaikin-edges}
\begin{array}{llcl}
\omega : & \lefteqn{P(v_1) E P(v_2) E \ldots  P(v_m) E} \\
p_1: & P(v_l) < E > P(v_r) & \rightarrow & 
	P(\frac{3}{4} v_l + \frac{1}{4}v_r) E P(\frac{1}{4} v_l + \frac{3}{4}v_r) \\
p_2 :  & \hspace{1.2cm} P(v) & \rightarrow & E
\end{array}
\end{equation}
A point $P(v)$ carries all the information needed to visualize it as, for example, a small circle centered at  $v$.  In contrast, the visualization of the edge $E$ between points $P(v_l)$ and $P(v_r)$ requires an interpretation rule~\cite{Karwowski2003dai,Kurth1994ggi} that is applied at the end of the derivation and gathers the information about the line's endpoints:
\begin{equation}
\label{eq:edge-interpretation}
\begin{array}{llcl}
h_E: & P(v_l) < E > P(v_r) & \rightarrow & L(v_l, v_r)
\end{array}
\end{equation} 
Module $L$ now carries all the information needed to draw a line from $v_l$ to $v_r$.  We assume that production~\ref{eq:edge-interpretation} complements all L-systems using explicit edge representation (module $E$).

A variant of L-system~\ref{eq:Chaikin-edges} is given below:
\begin{equation}
\label{eq:B-spline}
\begin{array}{lccl}
\omega :  & \hspace{-3.1cm} \lefteqn{P(v_1) E P(v_2) E \ldots  P(v_m) E} \\
p: & P(v_l) < E > P(v_r)   & \rightarrow & E P(\frac{1}{2} v_l + \frac{1}{2} v_r) E  \\
q_1: & P(v_l) < E > P(v_r) & \rightarrow & P \left( \frac{1}{2} v_l + \frac{1}{2} v_r \right)  \\
q_2: & P(v) & \rightarrow & E
\end{array}
\end{equation}
\begin{figure}
\centering
\includegraphics[width=1\linewidth]{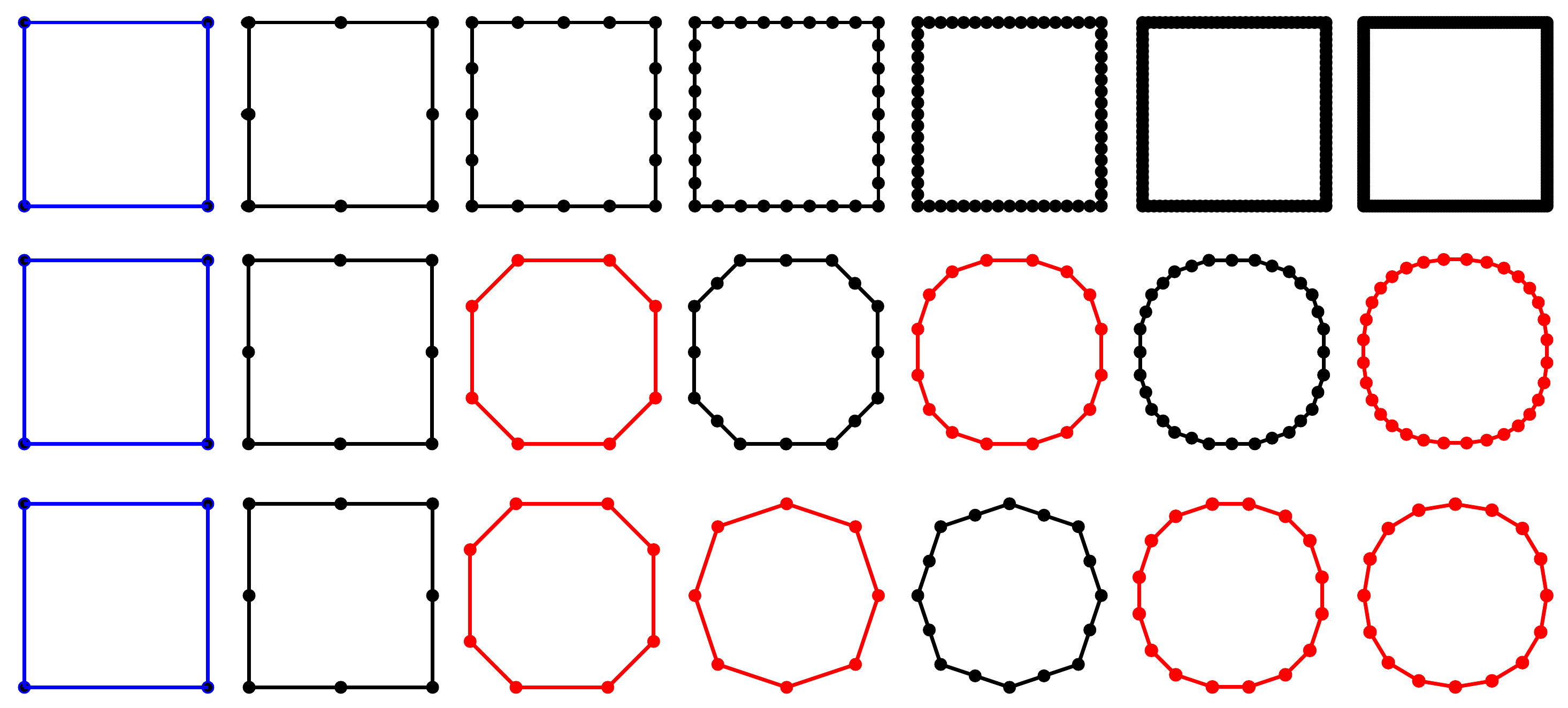}
\caption{Illustration of the Lane-Riesenfeld algorithm for generating B-splines of arbitrary degree $n+1$ (L-system~\protect\ref{eq:B-spline}).  Beginning with the initial polygon (shown in blue), the algorithm proceeds in cycles consisting of a single derivation step using production $p$ (result shown in black) followed by $n \geq 0$ steps using productions $q_1$ and $q_2$ (results shown in red).  Derivations of a linear ($n=0$), quadratic ($n=1$) and cubic B-spline ($n=2$) are shown in consecutive rows.}
\label{fig:B-spline}
\end{figure}

\noindent
The set of productions is partitioned here into two subsets, labeled $p$ and $q$.  Production $p$ inserts a new vertex in the middle of each edge, thus subdividing it into two halves. Productions $q_1$ and $q_2$ replace the predecessor polygon with a new polygon that has all vertices placed at the midpoints of the predecessor's edges.   Derivation proceeds in cycles.  Each cycle consists of a single application of production $p$ followed by $n \geq 0$  applications of productions $q_1$ and $q_2$ (Figure~\ref{fig:B-spline}).  For $n=1$, such a cycle produces the same result as one derivation step in L-system~\ref{eq:Chaikin-edges} (compare the middle row in Figure~\ref{fig:B-spline} with Figure~\ref{fig:Chaikin}).  Lane and Riesenfeld have shown that the polygons generated by L-system~\ref{eq:B-spline} converge to (uniform) B-spline curves of degree $n+1$~\cite{Lane1980atd} (see also~\cite{Maillot2001aus,Stam2001oss}). B-splines are widely used in geometric modeling due to their well understood geometric properties, and the relative ease with which diverse curves can be defined by interactively positioning the control points. L-system~\ref{eq:B-spline} provides a very concise description of this class of curves.  
 
\section{The de Casteljau algorithm}  

\begin{figure}
\hspace{0cm} a \hspace{3.68cm} b \hspace{3.6cm} c \hspace{3.65cm} d
\begin{center}
\includegraphics[width=1\linewidth]{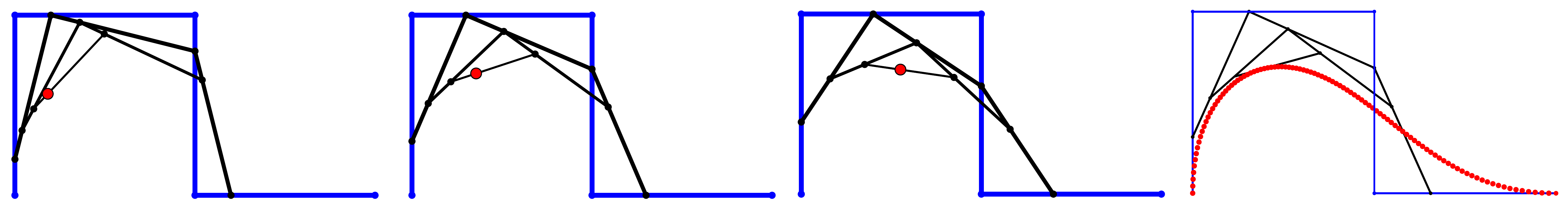}
\end{center}
\caption{Illustration of the de Casteljau algorithm (L-system~\protect\ref{eq:deCasteljau-intervals}).  a-c) Points obtained for parameter values $t = 0.2, 0.3$ and $0.4$, respectively.  The initial (open) control polygon is shown in blue, the polygons obtained in subsequent derivation steps are black, and the resulting point is shown in red.  d) The quartic B\'ezier curve obtained as a locus of points generated by the de Casteljau algorithm with the parameter $t$ increasing from 0 to 1 in increments $\Delta t = 0.01$.}
\label{fig:deCasteljau}
\end{figure}

The de Casteljau algorithm is considered ``probably the most important algorithm of all of computer-aided geometric design''~\cite[p. 32]{Farin2000teo}. Given an open control polygon $P(v^0_1) P(v^0_2) \ldots  P(v^0_n) $ with $n \ge 2$ vertices $P$ located at  $v^0_1,  v^0_2, \ldots , v^0_n$, the algorithm constructs a polygon  $P(v^1_1) P(v^1_2) \ldots  P(v^1_{n-1}) $ such that each point $P(v^1_i)$ subdivides the line segment $\overline{P(v^0_i) P(v^0_{i+1})}$ in proportion $t:1-t$.  The number $t$ is a parameter ranging between 0 and 1. This process is iterated $n-1$ times, ending with a single point $P(v^{n-1}_1)$.  The locus of points $P(v^{n-1}_1)$ obtained by the application of this algorithm for all $t \in [ 0, 1]$ is a curve, called the B\'ezier curve of degree $n-1$ defined by the control points (or polygon) $P(v^0_1) P(v^0_2) \ldots  P(v^0_n) $ (Figure~\ref{fig:deCasteljau}).  As with B-splines, the importance of B\'ezier curves stems from their well understood mathematical properties and the ease with which their shapes can be specified or changed by manipulating the control polygons.

The above process of finding a curve point corresponding to the parameter value $t$ can be concisely expressed using a simple L-system with axiom $\omega$ and productions $p_1, p_2$:
\begin{equation}
\label{eq:deCasteljau}
\begin{array}{llcl}
\omega : & \lefteqn{P(v_1) P(v_2) \ldots  P(v_n)} \\
p_1: & P(v) > P(v_r) & \rightarrow & P((1-t)v + t v_r) \\
p_2: & P(v)  & \rightarrow & \epsilon
\end{array}
\end{equation}
The first production replaces each point in the predecessor string with an affine combinations of this point and its neighbor to the right, thus capturing the essence of the de Casteljau algorithm.  The second production erases the last point in the sequence.  The same results are obtained by replacing the right-context-sensitive production $p_1$ with its left-context-sensitive counterpart:
\begin{equation}
\label{eq:deCasteljau-left}
\begin{array}{llcl}
p'_1: & P(v_l) < P(v) & \rightarrow & P((1-t)v_l + t v)
\end{array}
\end{equation} 

\begin{figure}
\hspace{0.7cm} a \hspace{6.9cm} b
\begin{center}
\vspace{-2mm}
\includegraphics[width=1.0\linewidth]{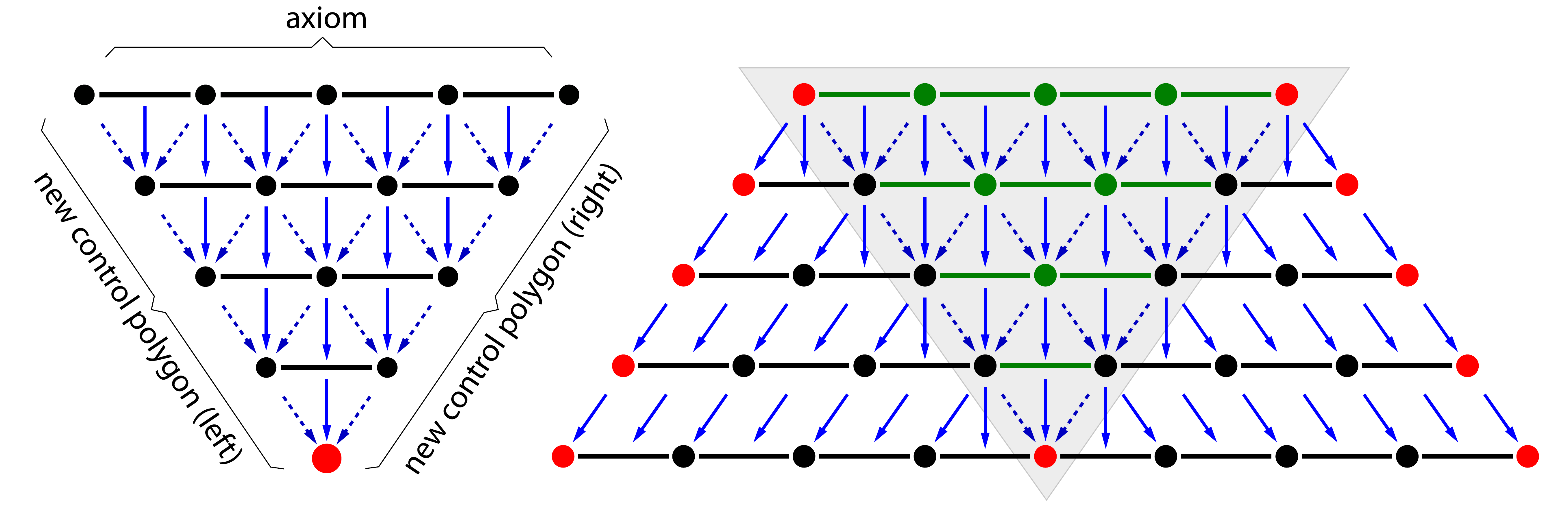}
\end{center}
\caption{(a) Example of the derivation tree generated by L-system~\protect\ref{eq:deCasteljau-intervals}, which implements the de Casteljau algorithm. A quartic B\'ezier curve ($n=5$) has been assumed. 
The L-system operates on polygons defined as sequences of vertices (circles) and edges (horizontal lines). The L-system axiom is given in the first row. Solid arrows relate the strict predecessor of each production to its successor. Dashed lines indicate the flow of contextual information specifying the location of each vertex.   The final result is a single vertex shown in red.   (b) Example of the derivation tree generated by L-system~\protect\ref{eq:deCasteljau-subdivision},  which aproximates B\'ezier curves of arbitrary degree using repetitive subdivision.  The first subdivision level of a quartic curve is shown.  The central part of the subdivision tree (within the shaded inverted triangle) implements the de Casteljau algorithm as shown in panel (a).  The left and right parts of the derivation tree propagate the vertices and edges established in the previous steps.  In order to operate properly, L-system~\protect\ref{eq:deCasteljau-subdivision} distinguishes between vertices and edges of different states and types.  For vertices, green denotes state $s=0$; black: $s=1$; red: $s=2$.  For edges, green denotes type $E$; black: type $I$.}
\label{fig:deCasteljau-derivation}
\end{figure}

\noindent
Unfortunately, neither production $p_1$ nor $p'_1$ captures the inherent symmetry of the de Casteljau construction.  As in the case of L-systems~\ref{eq:Chaikin-edges} and~\ref{eq:B-spline}, this shortcoming can be addressed by representing the control polygon as a one-dimensional cell complex: a sequence of points $P$ connected by edges $E$:
\begin{equation}
\label{eq:deCasteljau-intervals}
\begin{array}{llcl}
\omega :  & \lefteqn{P(v_1) E P(v_2) E \ldots  E P(v_n)} \\
p_1: & P(v_l) < E  > P(v_r) & \rightarrow & P((1-t)v_l + t v_r) \\
p_2: & E  <  P(v) >  E  & \rightarrow & E \\
p_3: & P(v)  & \rightarrow & \epsilon 
\end{array}
\end{equation}
Production $p_1$ now replaces an edge $E$ with a vertex $P$ dividing this edge in proportions $t: 1-t$.  Production $p_2$ performs a dual operation, replacing the vertex between two edges with an edge.  Finally, production $p_3$ erases the first and the last vertex of the predecessor polygon from the successor polygon.  The operation of this L-system is illustrated in Figure~\ref{fig:deCasteljau}, and the corresponding information flow if given by the derivation in Figure~\ref{fig:deCasteljau-derivation}a.

An alternative to the sequential evaluation of consecutive points approximating a B\'ezier curve can be described as follows.  Points at the inclined boundaries of the derivation tree (Figure~\ref{fig:deCasteljau-derivation}a) define two new control polygons with the same number of vertices as the original control polygon. It is known that B\'ezier curves defined by these new polygons subdivide the original curve into two segments~\cite{Lane1980atd} (see also ~\cite[page 34]{Farin2000teo}, for example).  Furthermore, the union of the new polygons is closer to the curve than the original polygon.  A B\'ezier curve can thus be generated by subdivision, i.e., by constructing two control polygons that define the same curve as the original polygon, and iterating this process until a desired accuracy of approximation of the curve by the union of its control polygons has been reached.  

The derivation tree for L-system~\ref{eq:deCasteljau-subdivision} implementing one subdivision level is shown in Figure~\ref{fig:deCasteljau-derivation}b.  The consecutively established vertices of the new control polygons propagate from one derivation step to the next until the resulting string is complete.   The algorithm builds new control polygons sequentially, by proceeding from the endpoints of the given control polygon inward.   The meeting point --- the final result of the de Casteljau algorithm --- is shared by both new polygons.  The L-system is given below:

% The above process can be specified and implemented using the following L-system:

\begin{equation}
\label{eq:deCasteljau-subdivision}
\begin{array}{llrl}
\omega:   & \lefteqn{P(v_1,2) E P(v_2,0) E \cdots  P(v_{n-1}, 0) E P(v_n,2)} \\
p_1: & P(v_l,s_l) < E  > P(v_r,s_r) & \rightarrow & P((1-t)v_l + t v_r, f(s_l,s_r)) \\
p_2: & E  <  P(v,s) >  E  &:  s = 0 \;  \rightarrow & E \\
p_3: & E  <  P(v,s) >  E  &:  s \neq 0 \; \rightarrow & I P(v,s) I \\
p_4: & P(v,s) >  E  &:  s \neq 0 \; \rightarrow & P(v,s) I  \\
p_5: & E  <  P(v,s)  &:  s \neq 0 \; \rightarrow & I P(v,s) \\
\end{array}
\end{equation}
We now explain the operation of this L-system in detail.  To produce the derivation tree in Figure~\ref{fig:deCasteljau-derivation}b, each point is characterized by its position $v$ (first parameter) and state $s$ (second parameter).  The states represent the following information:
\begin{enumerate}
\item[s=2:]   
an endpoint of a control polygon;
\item[s=1:]
a previously established interior vertex of the polygon being constructed;
\item[s=0:]
any other point.
\end{enumerate}
In addition, a distinction is made between edges $I$ that connect pairs of resultant points (including the endpoints) and edges $E$ that are still subject to changes by the algorithm.    The axiom $\omega$ specifies the initial control polygon, distinguishing between its endpoints and all other points.  Production $p_1$ and $p_2$ correspond to productions with the same labels in L-system~\ref{eq:deCasteljau-intervals} and capture the essence of the de Casteljau algorithm. A new element is function $f$ in production $p_1$, which assigns a state to the resultant point. The following rules apply to a vertex in state 0:
\begin{itemize}
\item
a vertex adjacent to an endpoint or one interior point becomes an interior point (s=1);
\item
a vertex adjacent to two interior points  becomes a new endpoint ($s=2$).
\end{itemize}
The remaining points retain their states.  With the assumed coding of states, function $f$ can be written as 
\begin{equation}
f(s_l,s_r) = \min(s_l,1) + \min(s_r,1).
\end{equation}
The assignment of states allows the L-system to distinguish between points and edges that are still subject of the Casteljau algorithm (productions $p_1$ and $p_2$) and points that are ready to be propagated to the resulting string (productions $p_3$ to $p_5$).  The propagation itself is effected by the identity productions assumed to operate on points and edges to which no other production applies.

To iterate the subdivision process, the states of points and the types of edges are re-initialized before the next subdivision cycle using productions:
\begin{equation}
\label{eq:deCasteljau-reinitialize}
\begin{array}{llll}
q_1: & P(v,s)  \; :  \;  s = 1 & \rightarrow & P(v,0) \\
q_2: & I  &  \rightarrow  & E
\end{array}
\end{equation}
\begin{figure}
\hspace{0.7mm} a \hspace{3.5cm} b \hspace{3.6cm} c \hspace{3.57cm} d
\begin{center}
\includegraphics[width=1\linewidth]{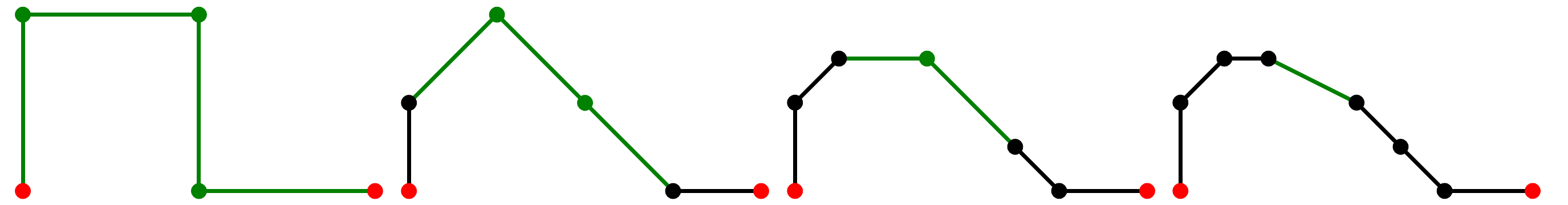}
\end{center}
\hspace{0.7mm} e \hspace{3.55cm} f \hspace{3.6cm} g \hspace{3.55cm} h
\begin{center}
\vspace{-1.5mm}
\includegraphics[width=1\linewidth]{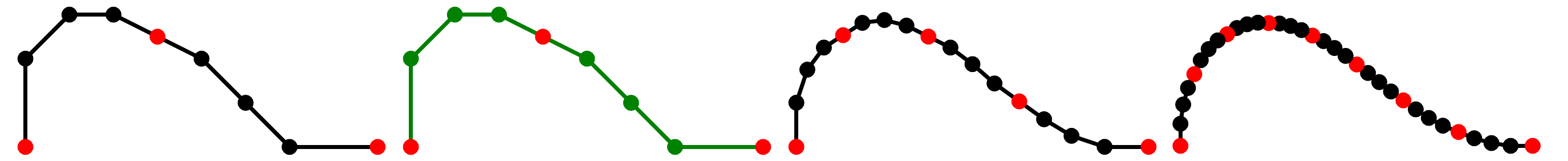}
\end{center}
\caption{Generation of a quartic B\'ezier curve with the de Casteljau subdivision algorithm. a) The initial polygon.  b-e) The first subdivision cycle using production set~\protect\ref{eq:deCasteljau-subdivision}. f) Re-initialization preceding the second subdivision cycle using production set~\ref{eq:deCasteljau-reinitialize}.  g-h) The result of the second and third cycle of subdivision. Vertex states and edge types are color-coded  as in Figure~\protect\ref{fig:deCasteljau-derivation}.}
\label{fig:deCasteljau-subdivision}
\end{figure}

\vspace{-5mm}
\noindent
Given an initial polygon with $n+1$ points, the L-system with productions~\ref{eq:deCasteljau-subdivision} and~\ref{eq:deCasteljau-reinitialize} thus generates a B\'ezier curve of degree $n$ by iterating the cycle of $n$ derivation steps using production set~\ref{eq:deCasteljau-subdivision}, followed by one step using production set~\ref{eq:deCasteljau-reinitialize}.  The parameter $t$ in production $p_1$ is typically set to 0.5 to achieve a relatively uniform distribution of points approximating the curve.  An example of a quartic B\'ezier curve generated by this process is shown in Figure~\ref{fig:deCasteljau-subdivision}.

\begin{figure}
\hspace{3.2cm} a \hspace{3.9cm} b
\begin{center}
\vspace{-3.5mm}
\includegraphics[width=0.6\linewidth]{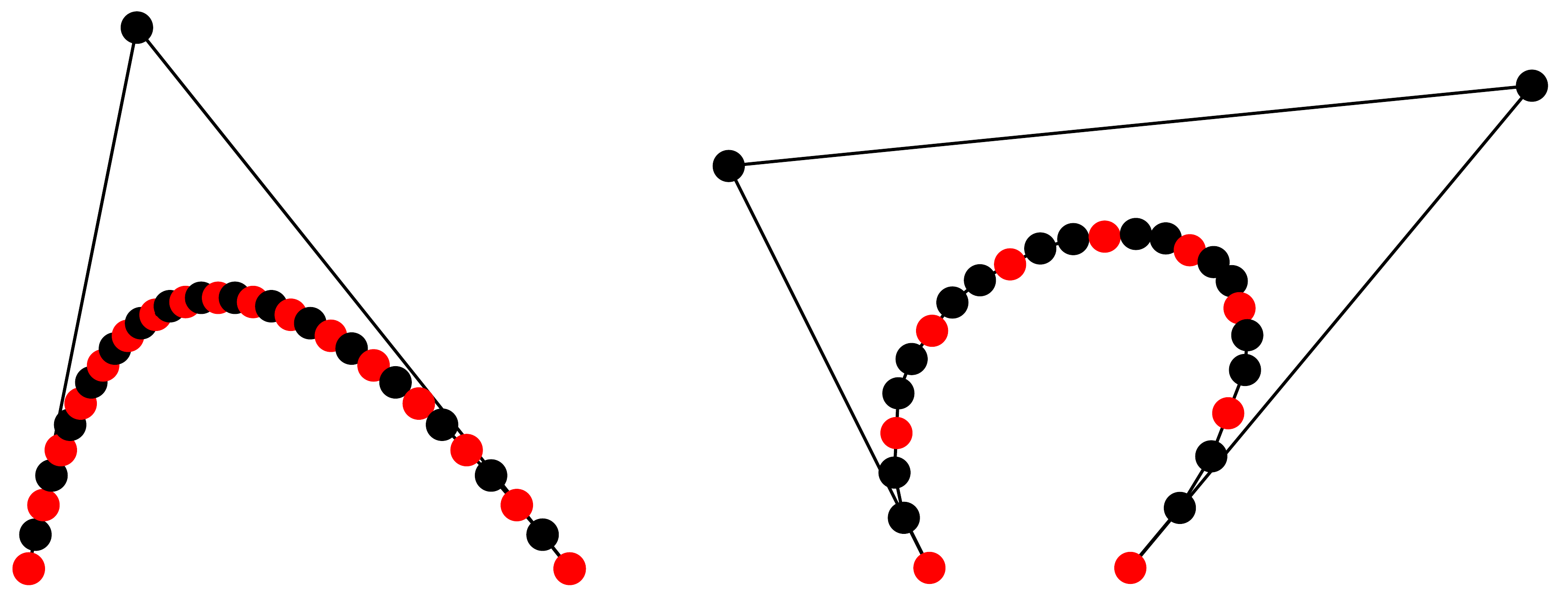}
\end{center}
\caption{Illustration of the de Casteljau subdivision algorithm, specialized to curves of a given degree.  a)  A quadratic B\'ezier curve generated using L-system~\protect\ref{eq:quadraticBezier}.  b) A cubic B\'ezier curve generated using L-system~\protect\ref{eq:cubicBezier} (or~\protect\ref{eq:cubicBezier-proper}). Points $P$ are shown in red, points $Q$ in black.}
\label{fig:Bezier-quad-cubic}
\end{figure}

If the curve degree is known in advance,  the result of a complete subdivision cycle applied to a set of internal points can be precomputed and encapsulated in a single production.  For example, the following L-system generates quadratic B\'ezier curves:
\begin{equation}
\label{eq:quadraticBezier}
\begin{array}{llll}
\omega:   & P(v_1) E Q(v_2) E P(v_3) \\
p: & P(v_l) E < Q(v) > E P(v_r)  & \rightarrow & 
	 Q(\frac{1}{2} v_l + \frac{1}{2} v)
	E P(\frac{1}{4} v_l + \frac{1}{2} v + \frac{1}{4} v_r) E 
	Q(\frac{1}{2} v + \frac{1}{2} v_r)
\end{array}
\end{equation}
\noindent
Here we distinguish points by their type ($P$ or $Q$) rather than state.  Points $P$ are the endpoints of a control polygon, and are not affected in the subsequent derivation steps.  Points $Q$ are the interior vertices and can be replaced with another set of vertices in the next derivation step.  Generation of a quadratic B\'ezier curve using L-system~\ref{eq:quadraticBezier} is illustrated in Figure~\ref{fig:Bezier-quad-cubic}a. 

The principle of this construction carries over to B\'ezier curves of degrees $n>2$.  However, the number  $n-1$ of the interior control points that need replacing is then greater than one.  Such a replacement can be effected most simply using a pseudo-L-system, in which multi-module strict predecessors can be rewritten at once~\cite{Prusinkiewicz1986}.  For example, a cubic B\'ezier curve (Figure~\ref{fig:Bezier-quad-cubic}b) is generated by the following pseudo-L-system:   
\begin{equation}
\label{eq:cubicBezier}
\begin{array}{llll}
\omega :  & P(v_1) E Q(v_2) E Q(v_3) E P(v_4) \\
p: & P(v_{ll}) E  < Q(v_l) E Q(v_r) > E P(v_{rr})   \rightarrow  \\
	& \hspace{1.8cm} Q(\frac{1}{2} v_{ll} + \frac{1}{2} v_l) E Q(\frac{1}{4} v_{ll} 
			+ \frac{1}{2} v_l + \frac{1}{4} v_r)  \\
	& \hspace{1.8cm} E P(\frac{1}{8} v_{ll} + \frac{3}{8} v_l + \frac{3}{8} v_r + \frac{1}{8} v_{rr}) E \\
	& \hspace{1.8cm} Q(\frac{1}{4} v_l + \frac{1}{2} v_r 
			+ \frac{1}{4} v_{rr}) E Q(\frac{1}{2} v_r + \frac{1}{2} v_{rr})
\end{array}
\end{equation}
An equivalent proper L-system can be obtained by dividing the predecessor and the successor of production $p$ into three parts, for example:
\begin{equation}
\label{eq:cubicBezier-proper}
\begin{array}{llll}
\omega :  & P(v_1) E Q(v_2) E Q(v_3) E P(v_4) \\
p_1: & P(v_{ll}) E  < Q(v_l) > E Q(v_r) E P(v_{rr})   & \rightarrow & 
	 Q(\frac{1}{2} v_{ll} + \frac{1}{2} v_l) E Q(\frac{1}{4} v_{ll} 
			+ \frac{1}{2} v_l + \frac{1}{4} v_r)  \\
p_2 : & P(v_{ll}) E  Q(v_l) < E > Q(v_r) E P(v_{rr})   & \rightarrow &
	 E P(\frac{1}{8}v_{ll} + \frac{3}{8}v_l + \frac{3}{8}v_r + \frac{1}{8}v_{rr}) E \\
p_3: & P(v_{ll}) E  Q(v_l) E < Q(v_r) > E P(v_{rr})   & \rightarrow &
	Q(\frac{1}{4} v_l + \frac{1}{2} v_r 
			+ \frac{1}{4} v_{rr}) E Q(\frac{1}{2} v_r + \frac{1}{2} v_{rr})
\end{array}
\end{equation}
Thus, L-systems succinctly express both the basic de Casteljau algorithm and extensions that generate B\'ezier curves of arbitrary or fixed degree using subdivision.

\section{Rational curves}

The algorithms discussed so far have been illustrated with examples of planar curves.  However, the assumption of planarity is not necessary, and the algorithms operate equally well in three and more dimensions.  This provides a means of generating curves in space, and also leads to a useful extension of B\'ezier and B-spline curves to their rational counterparts~\cite{Farin2000teo,Rockwood1996ica}.  

Rational curves are generated in a higher-dimensional space, then projected to lower dimensions  using a perspective projection.  Assuming $z \neq 0$, the projection of a 3D point $P(x,y,z)$ on the plane $z=1$ from the origin $O(0,0,0)$  of the underlying coordinate system is the 2D point $P'(\frac{x}{z}, \frac{y}{z})$. Identifying point locations with their coordinates, this projection can be accomplished by the interpretation rule
\begin{equation}
\label{eq:points-interpretation}
\begin{array}{llcl}
h_P: & P(x,y,z) & \rightarrow & P'(\frac{v_x}{v_z}, \frac{v_y}{v_z})
\end{array}
\end{equation} 
The subsequently applied edge interpretation rule (production~\ref{eq:edge-interpretation}) should operate on the projected points $P'$ rather than the original 3D points $P$, thus taking the form
\begin{equation}
\label{eq:edge-rational}
\begin{array}{llcl}
h_E: & P'(v'_l) < E > P'(v'_r) & \rightarrow & L(v_l, v_r)
\end{array}
\end{equation} 
 
\begin{figure}
\hspace{2.0cm} a \hspace{4.25cm} b \hspace{4.15cm} c
\begin{center}
\includegraphics[width=0.8\linewidth]{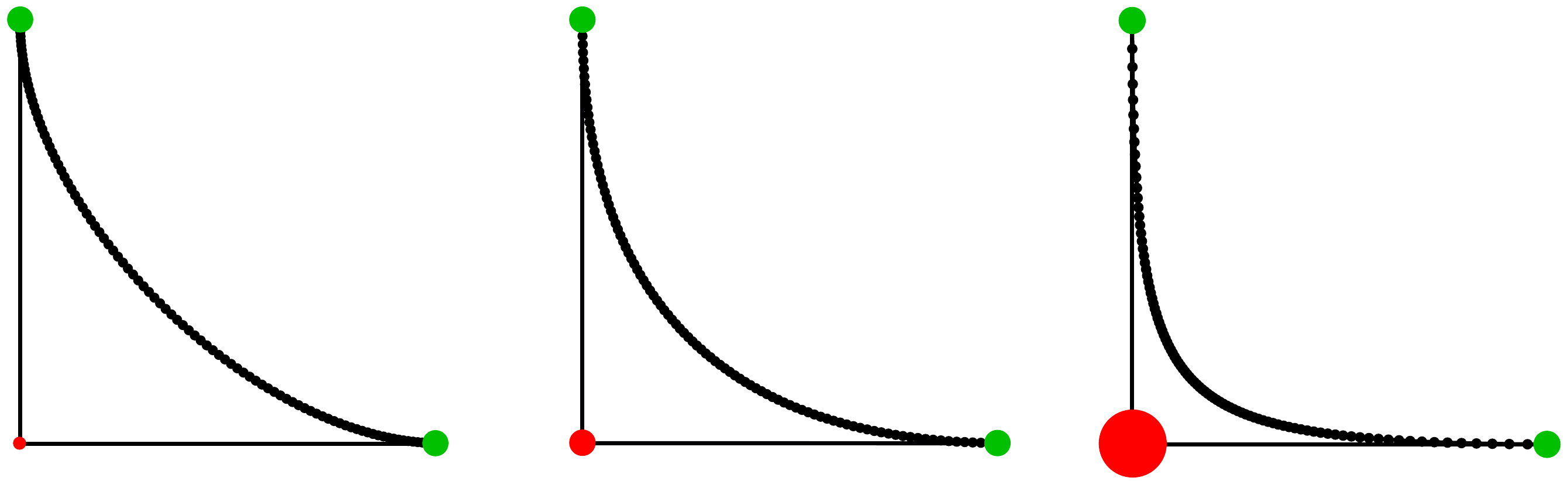}
\end{center}
\caption{Examples of rational quadratic B\'ezier curves generated using the de Casteljau algorithm operating in three dimensions. The endpoints of the control polygon (green) have weight 1.  The internal control point (red) has weight: a) 0.5, b) 1, and c) 2.5. The weights of the control points are indicated by their radius. Example based on~\protect\cite[pp. 124--125]{Rockwood1996ica}.}
\label{fig:Bezier-rational}
\end{figure}

 A 2D point $P'(v_x, v_y)$ can thus be represented by different 3D points of the form $P(wv_x, wv_y, w)$.  The parameter $w$ is called the weight of point $P'$.  Weights of control points do not affect the shape of the control polygon, but bias the resulting curve towards the points with a higher weight. This provides an additional means of controlling the shape of rational curves, beyond the manipulation of the control polygon.  Examples of quadratic rational B\'ezier curves generated with the de Casteljau algorithm (L-system~\ref{eq:deCasteljau-intervals}) using the same control polygon, but different point weights, are presented in Figure~\ref{fig:Bezier-rational}.  

\section{Conclusions}

We have shown that several fundamental algorithms for the geometric modeling of curves can be succinctly expressed using L-systems with affine geometry interpretation.  Examples include the Lane-Riesenfeld algorithm for generating uniform B-splines of arbitrary degree, the de Casteljau algorithm 
for generating B«ezier curves, and their extensions to rational curves. Our results generalize the previously reported geometric-modeling applications of L-systems, which were limited to subdivision curves~\cite{Prusinkiewicz2003lsd}.  Both the previous and current results demonstrate that L-system specifications closely match verbal descriptions of the modeling algorithms, and capture the information flow underlying their operation. This narrows the semantic gap between intuition and mathematical formalism, making L-systems useful as a notation for presenting the algorithms.  These advantages of L-systems are similar to those observed in plant modeling, and stem from the same features.  The key feature is the ease of expressing geometric algorithms that operate locally on structures with a varying number of components.  This ease is achieved through index-free notation that emphasizes the topological relations between components (c.f. Section~\ref{sec:Introduction}).  Whether the use of L-systems could also lead to a notational  and conceptual simplification of the theorem proofs considered in geometric modeling remains an interesting open question.

\section{Acknowledgments}

All examples presented in this paper were implemented using the L-system-based modeling program {\tt lpfg} developed by Radoslaw Karwowski~\cite{Karwowski2003dai}.  The correspondence between the mathematical notation used in this paper and the format of L-system specifications in {\tt lpfg} is discussed in~\cite{Prusinkiewicz2003lsd}. We thank Thomas Burt for extending {\tt lpfg} with context-sensitive interpretation rules, and Adam Runions for insightful comments on the draft manuscript.  The support of this research by the Natural Sciences and Engineering Research Council of Canada Discovery Grants to P.P. and F.S. is gratefully acknowledged.

%\newpage

\bibliographystyle{eptcs}
\bibliography{../BIB/l-systems,../BIB/graphics,../BIB/fractals}

\end{document}